\documentclass[conference]{IEEEtran}

\usepackage{graphicx}
\usepackage{svg}
\usepackage{subcaption}
\usepackage{hyperref} 
\usepackage{amsmath} 
\usepackage{amssymb} 
\usepackage{fancyhdr} 
\usepackage{pifont} 
\usepackage{xcolor}
\usepackage{soul}
\usepackage{adjustbox}
\usepackage{pgfgantt}
\usepackage{cite}
\ifCLASSINFOpdf
\else
\fi
\begin{document}
%
\title{Communication Characterization of AI Workloads for Large-scale Multi-chiplet Accelerators}
%
%
%


\author{
    Mariam Musavi, Emmanuel Irabor, Abhijit Das, Eduard Alarcón, and Sergi Abadal\\
    NaNoNetworking Center in Catalunya (N3Cat), 
    Universitat Polit\`ecnica de Catalunya (UPC), Barcelona, Spain
}

\markboth{Journal of \LaTeX\ Class Files,~Vol.~14, No.~8, August~2015}%
{Shell \MakeLowercase{\textit{et al.}}: Bare Demo of IEEEtran.cls for IEEE Journals}
%



\maketitle

\begin{abstract}
Next-generation artificial intelligence (AI) workloads are posing challenges of scalability and robustness in terms of execution time due to their intrinsic evolving data-intensive characteristics. In this paper, we aim to analyse the potential bottlenecks caused due to data movement characteristics of AI workloads on scale-out accelerator architectures composed of multiple chiplets. Our methodology captures the unicast and multicast communication traffic of a set of AI workloads and assesses aspects such as the time spent in such communications and the amount of multicast messages as a function of the number of employed chiplets. Our studies reveal that some AI workloads are potentially vulnerable to the dominant effects of communication, especially multicast traffic, which can become a performance bottleneck and limit their scalability. Workload profiling insights suggest to architect a flexible interconnect solution at chiplet level in order to improve the performance, efficiency and scalability of next-generation AI accelerators.
\end{abstract}


%
\IEEEpeerreviewmaketitle

\section{Introduction}\label{sec:introduction-label}
Artificial Intelligence (AI) applications have revolutionized multiple fields such as natural language processing, genomics, medical and health systems, graph analytics, and data analytics, and others \cite{Jouppi2021} \cite{gu2023gendp} \cite{Dadu2021}. However, the impressive feats that AI can achieve are often accompanied by very intense computational demands, which are saturating the boundaries of status-quo computing infrastructures \cite{wu2022sustainable}. \let\thefootnote\relax\footnotetext{Authors gratefully acknowledge funding from the European Commission through projects with GA 101042080 (WINC) and 101189474 (EWiC) and from Generalitat de Catalunya through the ICREA Academia Award 2024.} 

To address this, specialized hardware (HW) accelerators are designed to target specific computational-intensive AI workloads in an efficient manner. GPUs can be considered as hardware accelerators for graphics, that have been adapted to also better support AI workloads. However, the past decade has seen the emergence of AI accelerators with tensor or transformer cores, which cater to the specific computational needs of AI \cite{Garg2024}. Generally, the architecture of AI accelerators is comprised of an off-chip memory, a shared on-chip memory known as global buffer, and an array of processing elements (PEs) connected via a Network-on-Chip (NoC) \cite{chatarasi2021marvel}. 


As modern AI models are evolving in size and diversity to accommodate multiple applications \cite{odema2024scar} or to implement input-dependent models such as Graph Neural Networks (GNNs) \cite{garg2022understanding}, versatile and high-performance accelerators are required to support their execution. This can be achieved with reconfigurable dataflows in FPGAs \cite{hwang2020centaur}, flexible accelerator architectures \cite{Kao2022, kwon2018maeri}, or by resorting to less efficient but more general-purpose GPU and CPU architectures \cite{li2022ai}.

\begin{figure}
  \centering
  \adjustbox{trim=1cm 0.7cm 1cm 0.7cm}{
  \includegraphics[width=1\linewidth]{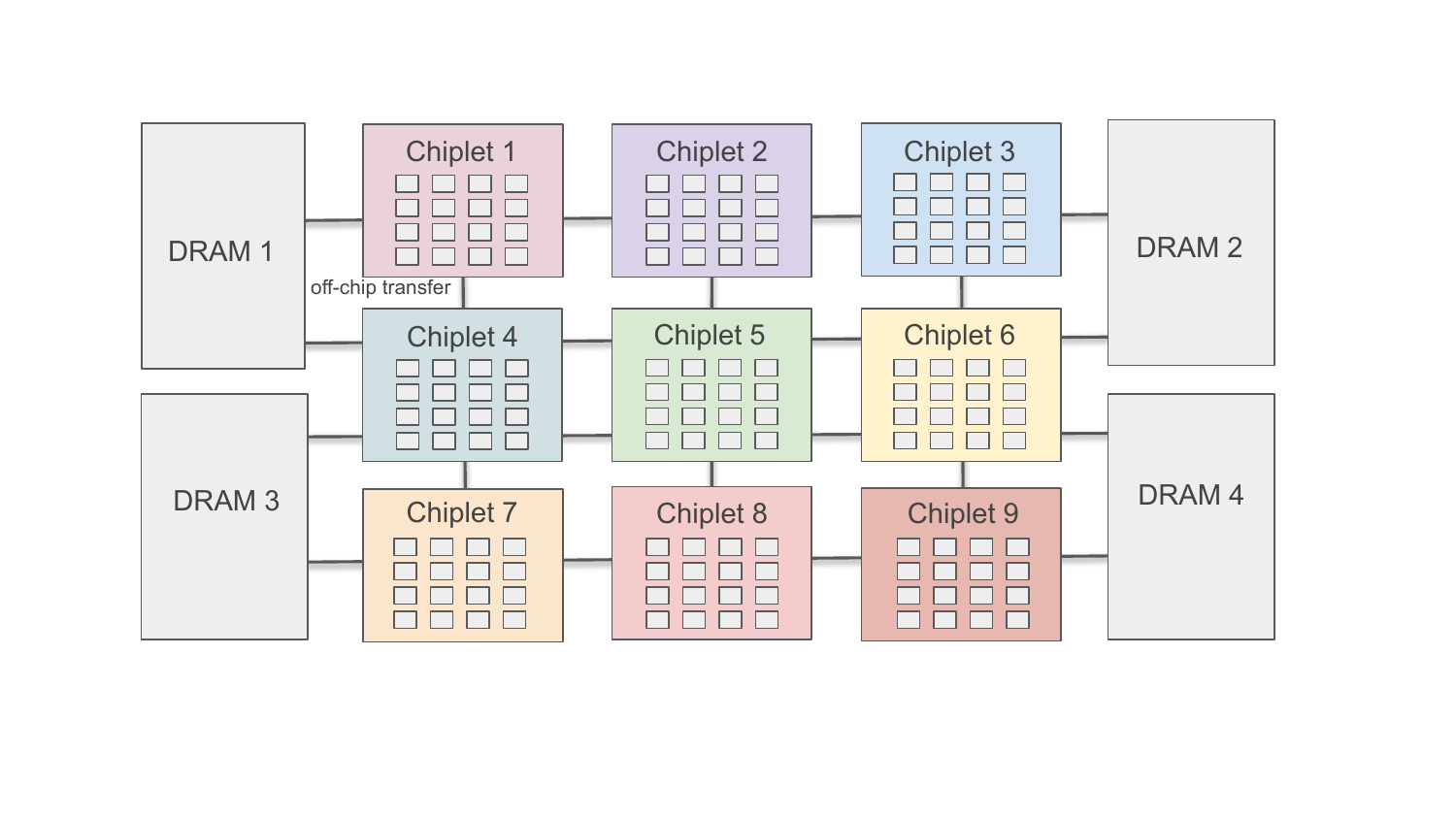}
    }\vspace{-0.4cm}
  \caption{An illustration of multi-chiplet architecture with 3x3 computing chiplets and 4 DRAM chiplets.}\vspace{-0.5cm}
  \label{fig:Chiplet-label}
\end{figure}

Another alternative compatible with the scaling and adaptation of AI accelerators is the use of chiplets. Indeed, chiplet technology is a promising enabler that provides a way to scale AI accelerators, by combining together multiple specialized (and potentially heterogeneous) AI accelerator chiplets in a single computing platform, as illustrated in Fig. \ref{fig:Chiplet-label}. These chiplets are interconnected among themselves and to memory via on-package links, typically through silicon interposers or organic substrates, in order to create a Network-on-Package (NoP) \cite{beck2018zeppelin, kannan2015enabling, vivet20202}. This has been proposed in multiple works, including SIMBA \cite{shao2019simba} or WIENNA \cite{guirado2021dataflow}, among others.

It is worth noting that AI accelerator chiplets can spend more than 90 percent of the total system energy on memory-bound tasks, to fetch data, as illustrated in \cite{boroumand2018google}. This is due to not only the limited speed of memory modules, but more importantly, the relatively slow chiplet-to-chiplet data transfers, which can often dominate the computation energy due to traversing long interconnects. This issue is exacerbated by the need to use multicast communication in many of the dataflows that are used in AI accelerators \cite{guirado2021dataflow, sze2017efficient}. 


Despite the importance of chiplet-level communication in multi-chip AI accelerators, the communication traffic within such systems has not been explored in depth nor characterized. Traffic characterization and modeling has been well studied in CPU and GPU platforms for a variety of workloads \cite{soteriou2006statistical, abadal2015multicast, lal2014gpgpu}, yet a similar analysis is missing in the proposed context.


The main contribution of this work is the profiling of workload-specific data movements across a set of popular AI workloads in multi-chiplet AI accelerators under increasing size. We augment GEMINI \cite{cai2024gemini} to register communication packets and then parse the resulting traces to analyze their characteristics, for instance, the number of messages, destinations per multicast, or the number of NoP hops per message.





The remainder of the paper is structured as follows. Section~\ref{sec:relatedWork-label} presents the state of the art in multi-chip AI accelerators and workload characterization. Our methodology is presented in Section \ref{sec:characterisationMethodology-label}. Section \ref{sec:characterisationResults-label} presents the results obtained, followed by the conclusion and future work in Section \ref{sec:conclusion-label}. 

\section{Related Works}\label{sec:relatedWork-label}
\subsection{Scheduling and Mapping for AI Accelerators}
In the pioneering work of \cite{shao2019simba}, authors developed and characterized a multi-chip AI inference accelerator with a scalable and hierarchical interconnect architecture on a package, called SIMBA. The objective was to enhance the energy efficiency and reduce accelerator's inference latency by partitioning the non-uniform workload, considering communication-aware data placement, and implementing cross-layer pipelining. The mapper used in this work is Timeloop \cite{parashar2019timeloop}, and the cost of the HW architecture is estimated with Accelergy \cite{wu2019accelergy}. 

The authors of \cite{cai2024gemini} developed and characterized a multi-chip AI inference accelerator design framework, called GEMINI, that explores the design space to deliver architectures that, for a particular workload, minimize monetary cost and Energy-Delay Product (EDP). The mapper employed in this work is inter-layer pipelining using SET \cite{cai2023inter}, and a customized cost model was used to evaluate the cost of the architecture. These improvements significantly enhanced the overall system performance compared to the baseline mapping variants. 

In recent work, the authors of \cite{odema2024scar} developed a multi-chiplet-based multi-model AI inference accelerator that is scalable under heterogeneous traffic (i.e., data center and AR/VR) models with the objective to minimize EDP. The mapper used in this work is based on SET \cite{cai2023inter}, and the hybrid cost model is customized using MAESTRO \cite{kwon2020maestro}. 

In all these works discussed above and similar ones like MOHaM~\cite{das2024multi}, MAGMA~\cite{kao2022magma}, Herald~\cite{kwon2021heterogeneous}, etc., communication is either coarsely classified depending on the chosen dataflow or indirectly used as a metric for optimization of mapping. However, none of the previous works have conducted an in-depth characterization of the communication workload across AI algorithms and architectures.

\subsection{Workload Parallelism}
Work that considered to determine spatial and temporal tiling factors such as, PE dimensions, accumulator SRAM sizing, scratchpad SRAM sizing, and finds the best compute/buffer sizes using a mapping-first approach includes Timeloop \cite{parashar2019timeloop}, Sparseloop \cite{wu2022sparseloop}, COSA \cite{huang2021cosa} and DOSA \cite{hong2023dosa} to reduce memory access overheads and data replication. Works that exploit both intra- and layer loop ordering using inter-layer pipelining include TANGRAM \cite{gao2019tangram}, SET \cite{cai2023inter}, GEMINI \cite{cai2024gemini}, and most recently SCAR \cite{odema2024scar} to improve data re-usability reduce pipelining delays. Again, in these cases, communication cost is used directly or indirectly as a metric for the optimization of the mapping, but without providing a proper characterization of the different workloads.

\subsection{Communication Workload Analysis}
In addition, there are a few existing works that characterize communication in CPU workloads based on cycle-accurate simulation, obtaining spatial-temporal distributions or correlations between traffic flows, among others, for unicast and multicast communication \cite{soteriou2006statistical, abadal2015multicast, abadal2016characterization}. Similar investigations have been conducted on GPU architectures \cite{lal2014gpgpu}. Undoubtedly, all the above work outperformed the performance on workloads-chiplets mapping; however, there is still a gap left to fulfill characterizing workloads from inter-chiplet multicast communication perspective. This characterization has significant potential to assist mappers/schedulers to efficiently orchestrate computing resources and tailor data movement strategies according to the underlying communication distance bottlenecks in executing the next-generation AI workloads. 
To the best of our knowledge, there is a critical need to characterize multicast data movement pattern in order to obtain the key metrics in the realm of diverse AI workloads-chiplet mappings under scalability (i.e., increasing the size of chiplet array configurations).
\section{Characterization Methodology}\label{sec:characterisationMethodology-label}
\begin{figure}[!t]
  \centering
  \vspace{-0.4cm}
  \includegraphics[width=1\linewidth]{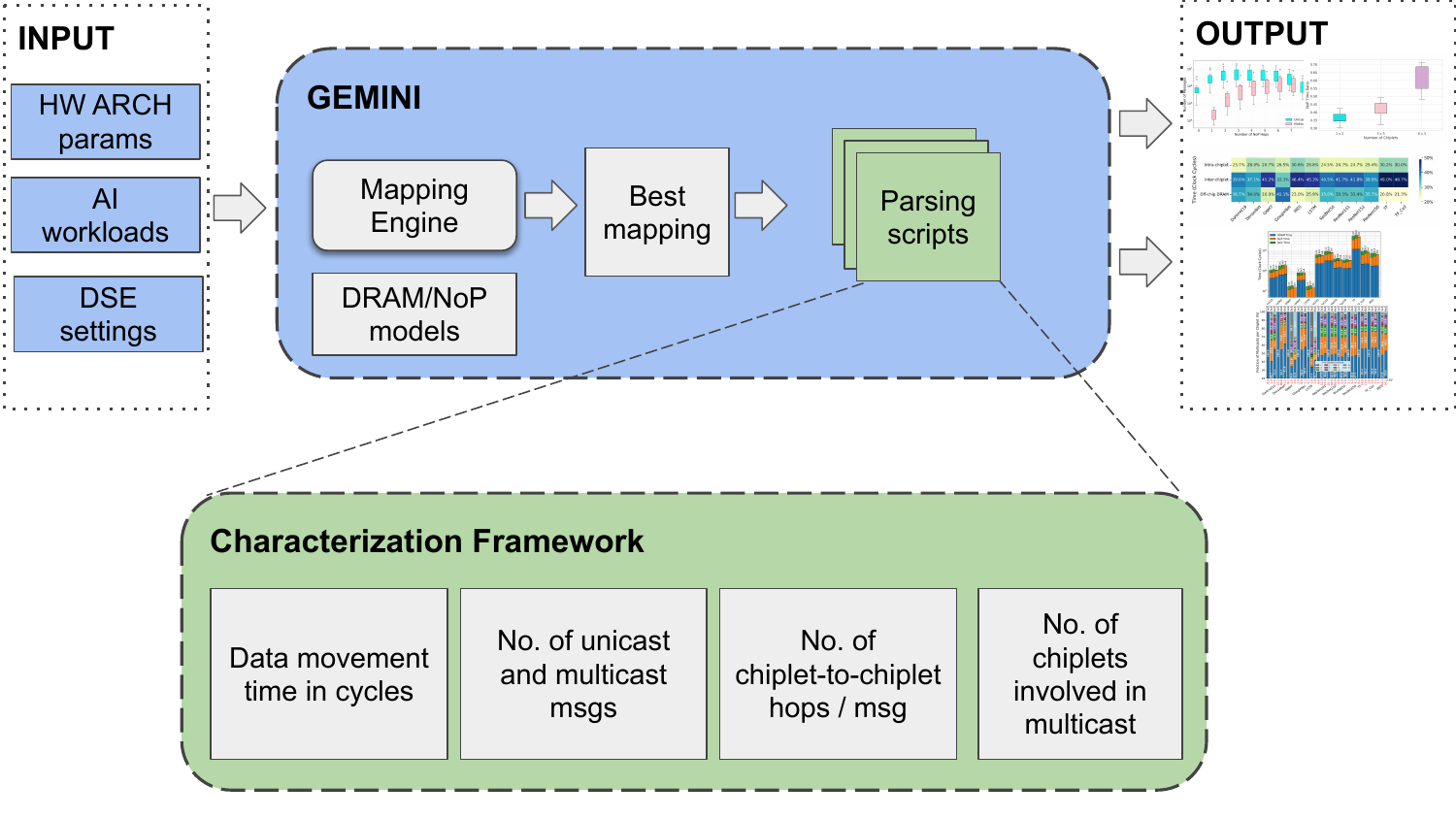}
  \caption{Methodology for characterizing data movement of AI workloads in multi-chip architectures, based on GEMINI \cite{cai2024gemini}.}
  \label{fig:methodology-label}
  \vspace{-0.5cm}
\end{figure}

The main objective of this work is to perform a traffic analysis to obtain communication behaviors of various AI workloads, in order to assess their scalability. To this end, we employ the methodology outlined in Fig. \ref{fig:methodology-label}, to capture important communication-related characteristics (e.g., multicast data movement) of all benchmark AI workloads. 
The communication workloads to be analyzed depend on multiple factors such as AI workload itself, the size of architecture, or how the workload is mapped into the architecture. Mapping requires decisions on the loop execution order and also regarding dataflow dependencies. The former determines the order in which the layer is mapped and executed on architecture, while later determines the inter-dependency of data provisioning between layers (i.e., the movement of output activation and partial sums, between preceding and subsequent layers). 

To abstract away from these decisions, here we use GEMINI \cite{cai2024gemini} to determine the best mapping for a particular combination of AI workload and multi-chip accelerator configuration. We wrote C++ functional scripts to obtain a set of communication traces from GEMINI, which are later parsed with the aim to extract profiles, which can further be processed and visualized using Python 3.10. We mainly focused on capturing multicast communication metrics, due to the severity of their effects when scaled, across all workloads and configurations. In particular, the communication metrics extracted are:
\begin{itemize}
    \item Amount of time in clock cycles spent performing the data movement by the relevant compute/memory chiplets.
    \item Number of unicast and multicast messages. 
    \item Number of chiplet-to-chiplet (NoP) hops per message.
    \item Number of chiplets involved in a multicast message.
\end{itemize}

Table \ref{tab:simulationParameters-label} presents the parameters explored in the mapping of $12$ AI inference workloads on the best HW architecture parameters obtained from GEMINI. We choose benchmark models that contain multi-branch classic residual, e.g. ResNet50, ResNet152, GoogleNet, Transformer (TF), TF Cell, and inception (iRES) structures with more intricate dependencies which are prevalent and widely used in various scenarios such as image classification and language processing. The bandwidths in the different communication levels, as well as the three multi-chiplet arrangements from $2$ to $18$ chiplets, are chosen based on GEMINI baselines. Regardless of the number of AI compute chiplets, the number of DRAM chiplets is set to four.

\begin{table}[t]
  \centering
  \caption{Simulation Parameters}\vspace{-0.1cm}
  \begin{tabular}{c|c}
    \hline
    Number of chiplets & (1x2), (3x3), (3x6) \\
    \hline
    NoP & XY Mesh, 4 GB/s/side  \\
    \hline
    NoC & XY Mesh, 8 GB/s/port \\
    \hline
    DRAM & 4 Chiplets, 16 GB/s/chiplet \\
    \hline    
    & Darknet19, DenseNet, GNMT, GoogleNet, \\
    AI Workloads & LSTM, ResNet101, ResNet101, ResNet152,\\
    & ResNet50, ResNext50, TF, TF Cell, iRES\\
    \hline 
  \end{tabular}  \label{tab:simulationParameters-label}
\end{table}
\section{Workload Characterization Results} \label{sec:characterisationResults-label}
This section presents our characterization results and discusses the insights obtained for multiple AI workloads on different multi-chip accelerator configurations.

\subsection{Communication Source Analysis}
\begin{figure}[!t]
  \centering 
  \includegraphics[width=0.9\linewidth]{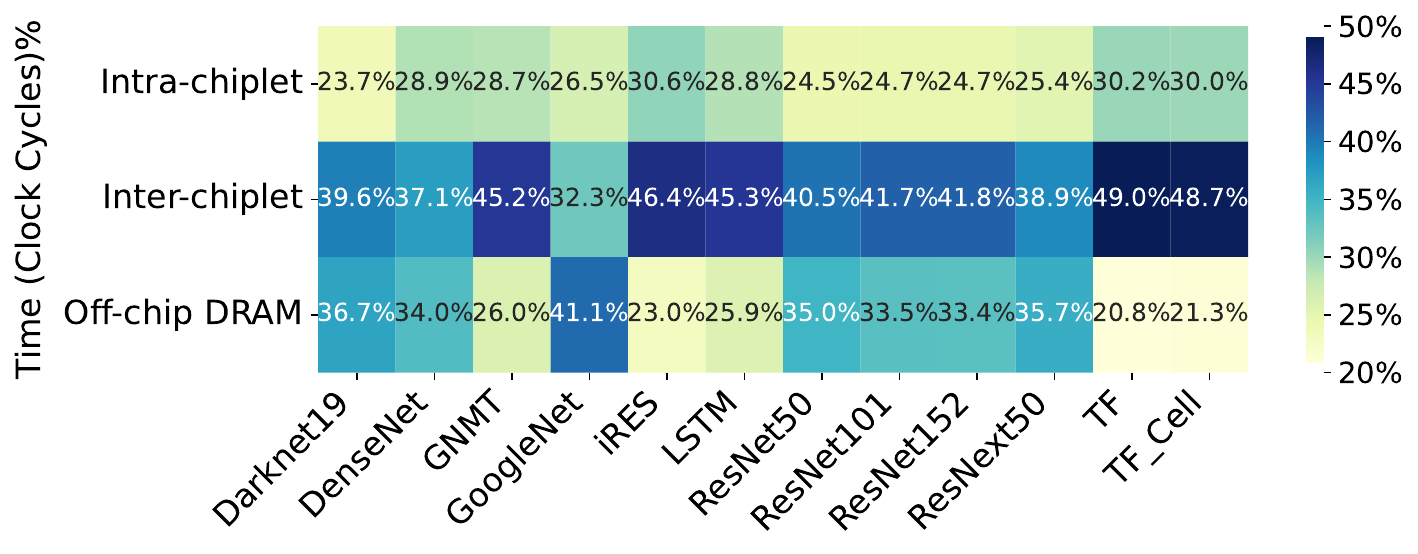}
  \caption{Fraction of overall execution time (in clock cycles) spent by on-chip, chip-to-chip and chip-to-DRAM data movement related tasks for each AI workload across configurations.}
  \label{fig:heatmap-label}
  \vspace{-0.5cm}
\end{figure}

Fig. \ref{fig:heatmap-label} reveals the sources of data movement by assessing the time spent by NoC, NoP, and DRAM across all workloads and averaged across system sizes. Notably, it can be observed that in most cases, the inter-chiplet data movement is the highest contribution among the three, especially for workloads such as iRES, LSTM, or TF.

\begin{figure}
  \centering 
  \includegraphics[width=1\linewidth]{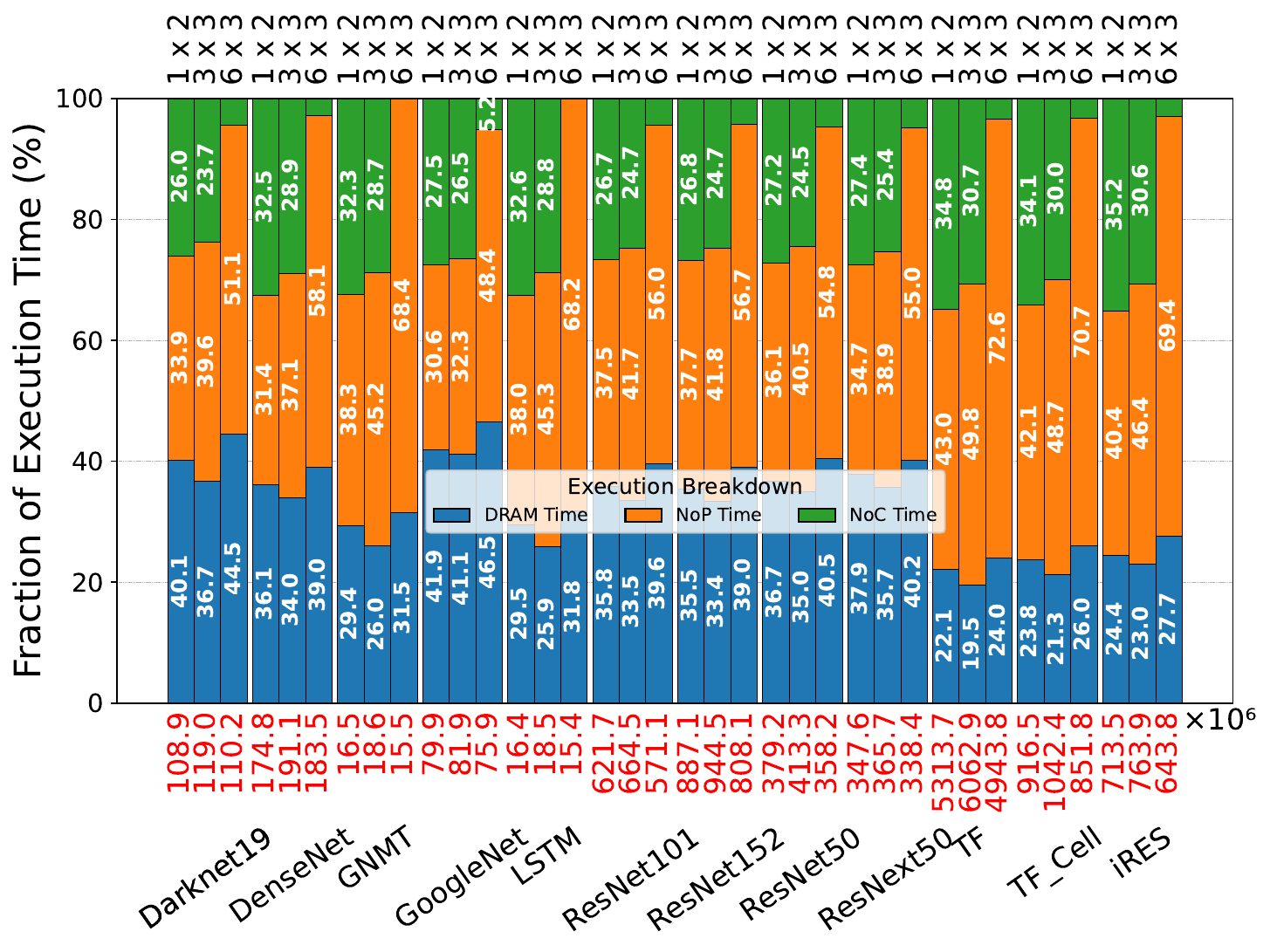}
  \caption{Fraction of data movement time spent in the DRAM, NoP and NoC across all AI workloads and chiplet array configurations. The total time in clock cycles is shown in red at the bottom of the plot.}
  \label{fig:allWorkloads-label}
  \vspace{-0.5cm}
\end{figure}

This aspect is further analyzed in Fig. \ref{fig:allWorkloads-label}, which shows the amount of time spent on data movement (i.e., communication time) contributed by each NoC, NoP and DRAM blocks for the different system sizes considered. During each workload execution, an increase in the number of chiplets directly contributes to an increase in NoP time, while indirectly impacting NoC time. As the number of chiplets grows, the number of computation cores also increases, enabling more efficient handling of computations.
However, it is evident that, under $6$x$3$ configuration, some AI workloads such as TF and TF\_Cell are heavily bottlenecked by the inter-chiplet communication time by $72.6$\% and $70.7$\%, respectively. In contrast, GoogleNet and DarkNet19 appear to be memory-bounded by $46.5$\% and $44.5$\%, respectively.

\begin{figure}[!t]
  \centering
  \includegraphics[width=0.8\linewidth]{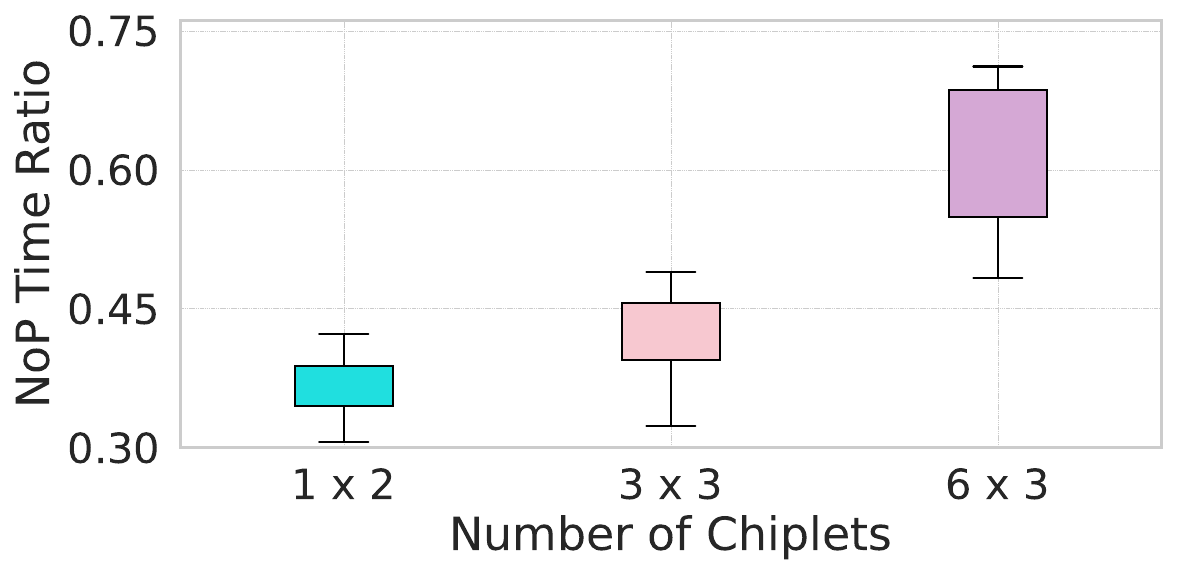}
  \caption{Fraction of time spent by NoP over the total communication time across all chiplet array configurations.}
  \label{fig:nopTimeRatio-label}
\end{figure}

In summary, Fig. \ref{fig:nopTimeRatio-label} shows in the form of a box plot, the fraction of data movement time spent by NoP over the sum of NoC, NoP and DRAM time for all workloads as a function of the number of chiplets in the system. It is evident from the plot that the amount of NoP time is directly proportional to the increase in network size, reaching values well above $50\%$ and even beyond $70\%$, which will eventually throttle the execution of the AI workload. Therefore, it is necessary to improve and design flexible data movement-aware architectures to eliminate the communication bottleneck.

\subsection{Multicast Traffic Analysis}
\begin{figure}[!t]
  \centering
  \includegraphics[width=1\linewidth]{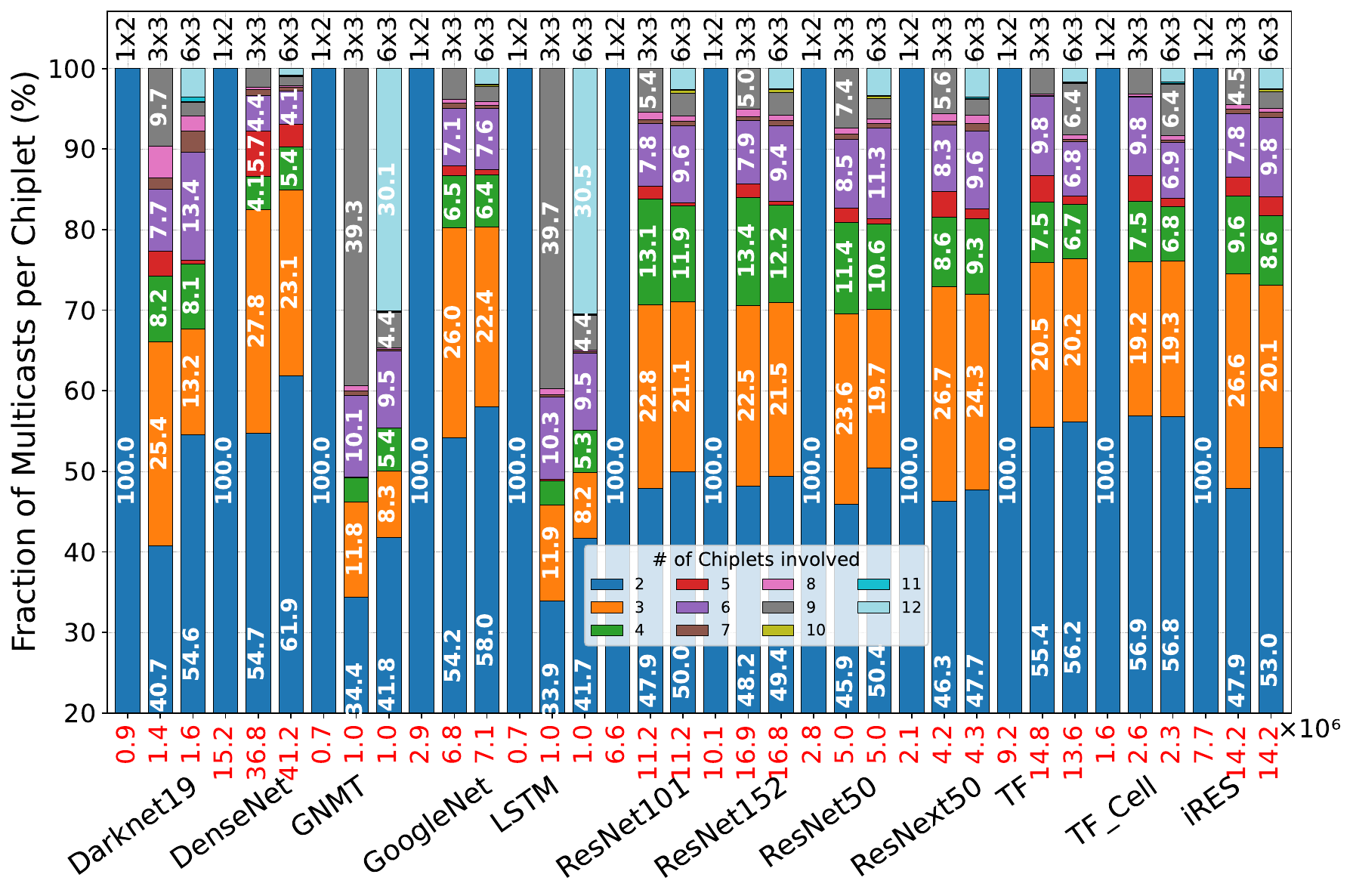}
  \caption{Number of multicast messages and number of destinations per message across all AI workloads and for multiple chiplet array configurations. The number of messages is shown in red at the bottom of the plot, whereas the number of destinations is shown in the form of a stacked bar where each color represents a given number of destinations.}
\label{fig:allWorkloadMsgs-label}
\vspace{-0.5cm}
\end{figure}

Fig. \ref{fig:allWorkloadMsgs-label} shows the fraction of multicast messages generated by each workload and the number of chiplets combination that are included in the destination set of each multicast message. First, it is evident from the plot that algorithms such as DenseNet ($15.2$, $36.8$ and $41.2$) $\times 10^6$ and TF ($9.2$, $14.8$ and $13.6$) $\times 10^6$ have a high number of multicast messages for 1x2, 3x3 and 6x3 chiplet configurations, respectively. Overall, it is observed that the importance of muilticasts with two, four, and six chiplet destinations is high. However, the number of multicast messages with the highest number of visited chiplets is very significant, i.e., $12$, especially in workloads such as LSTM ($30.1\%$) or GNMT ($30.5\%$).

\begin{figure}[!t]          
\centering 
\includegraphics[width=0.9\linewidth]{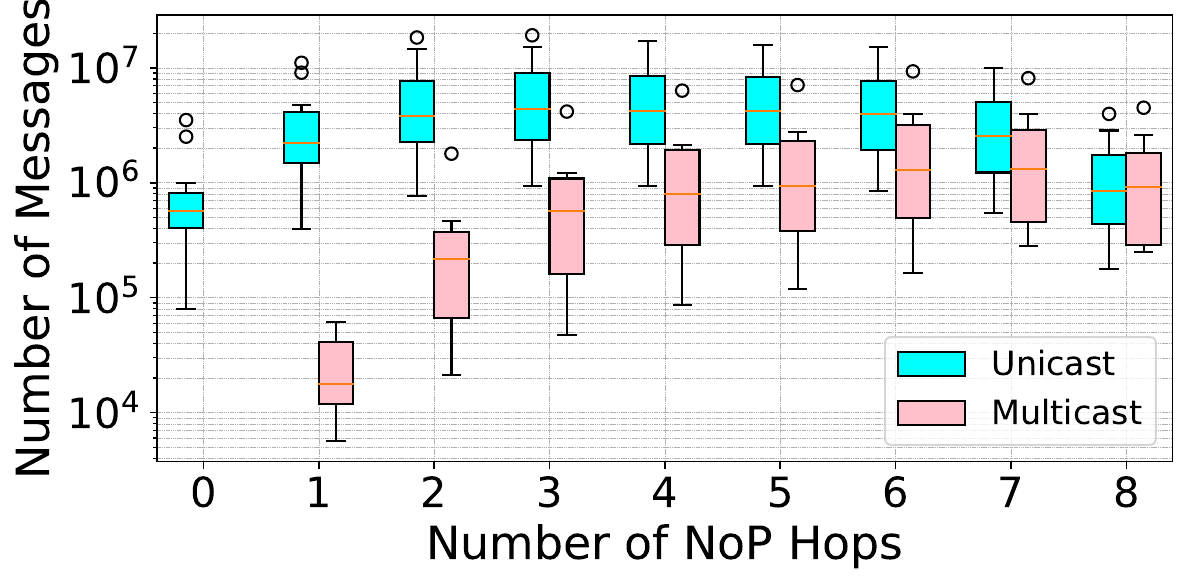}
\caption{Number of unicast and multicast messages across all AI workloads for the $6$x$3$ chiplet array configuration.}\label{fig:boxallWorkloadMsgs-label}
\vspace{-0.4cm}
\end{figure}

Finally, Fig. \ref{fig:boxallWorkloadMsgs-label} shows the distribution of unicast and multicast messages as a function of the number of NoP hops required in the $6$x$3$ chiplet array. This figure offers multiple insights. On one hand, the notion of spatial locality is not common in multi-chip AI accelerators, as the increasing number of messages with high number of NoP hops suggest. In contrast, even though the number of multicasts is significantly smaller than the number of unicasts, they are more prone to establish long-range connections. 

\subsection{Discussion}
In summary, workload-specific layers consist of data-dependent operators, with different compute, memory, and data movement requirements. As the number of chiplets increases, the number of communication hops also increases, which leads to a rise in the latency incurred due to moving data from memory to compute units. Hence, workload-specific multicast profiling is particularly useful to identify underlying communication-distance bottlenecks and to alleviate by improving the interconnect fabrics. 

Emerging interconnect technologies, i.e., wireless \cite{irabor2025, palesi2023wireless, guirado2021dataflow, abadal2022graphene} and/or optical-enabled interconnects \cite{taheri2024swint, steinman2023hummingbird}) are promising candidates to tackle the communication overhead in multi-chip accelerators~\cite{das2024chip}. An alternative is to reduce the communication intensity through novel memory and computing approaches such as in-memory computing (IMC), near-memory computing (NMC), or adopting vertical chip stacking technologies such as 3D-stacked memory.

Communication-wise, optical interconnects offer higher bandwidth and lower power consumption per transmitted bit. For instance, Hummingbird \cite{steinman2023hummingbird} employs an optical interconnect that allows all-to-all broadcast network with great benefits of reduced latency for long-distance data transfers, yet the scaling of such architecture remains unclear. Wireless technology, on the other hand, can become a powerful complement to wired interconnect fabrics offering dynamic bandwidth sharing, broadcast support, yet at the cost of reduced bandwidth. In this scheme, priority-based packet transmission of certain critical messages can be exploited to alleviate specific communication bottlenecks that happen if the workload is unbalanced within the wired and wireless planes of the multi-chip accelerator architecture. Hence, scaling-driven architecture, enabled by wireless communication or optical communication, has huge potential to address these challenging end-to-end latency constraints with high reliability and stability. 



\section{Conclusion and Future Work}\label{sec:conclusion-label}
We presented a multicast characterization tool, based on GEMINI \cite{cai2024gemini}, to register and analyze the communication workloads of AI algorithms mapped onto multi-chip AI accelerators. We observed that the communication latency has the potential to become an important bottleneck in most of the analyzed workloads, an aspect that is exacerbated as the communication intensity grows proportionally to the number of chiplets in the system. In particular, we identified multicast data movement as an important contributor to the overall required communication.
We posit that this analysis and modeling approach is critical to take into account the relevant trade-offs in designing the architecture for large-scale multi-chip AI accelerator systems. Specifically, we suggest addressing the communication bottlenecks by supplementing existing interconnects with emerging enablers such as wireless or optical interconnect technologies.







\bibliographystyle{IEEEtran}
\bibliography{IEEEabrv,./conf.bib}
%

%








\end{document}